\documentclass[aps,pra,preprintnumbers,showpacs,tightenlines]{revtex4}
\usepackage{amssymb}
\usepackage{amsmath}
\usepackage{graphicx}
\usepackage{epsfig}
\usepackage{subfigure}
\usepackage{amsfonts}
\usepackage{CJK}

\begin{document}

\title{Transferring multipartite entanglement among different cavities}

\author{Qi-Ping Su, Tong Liu, and Chui-Ping Yang$^{\star}$}

\address{Department of Physics, Hangzhou Normal University,
Hangzhou, Zhejiang 310036, China}

\address{$^\star$ yangcp@hznu.edu.cn}

\date{\today}

\begin{abstract} The transfer of quantum entanglement (or quantum coherence)
is not only fundamental in quantum mechanics but also important in
quantum information processing. We here propose a way to achieve the coherent transfer of $W$-class
entangled states of qubits among different cavities.
Because no photon is excited in each cavity, decoherence caused by the
photon decay is suppressed during the transfer. In addition, only one
coupler qubit and one operational step are needed and no classical pulses
are used in this proposal, thus the engineering complexity is much reduced
and the operation is greatly simplified. We further give a numerical
analysis, showing that high-fidelity transfer of a three-qubit $W$ state is
feasible within the present circuit QED technique. The proposal can be
applied to a wide range of physical implementation with various qubits such
as quantum dots, nitrogen-vacancy centers, atoms, and superconducting qubits.
\end{abstract}

 \maketitle
\date{\today}

\section{INTRODUCTION}

Quantum entanglement, as a cornerstone of quantum physics, plays an
important role in the foundation of quantum theory and has many potential
applications in quantum information processing (QIP) and communication. It
is known [1] that the two inequivalent and non-converted classes of
multipartite entangled states, i.e., Greenberger-Horne-Zeilinger (GHZ) [2]
state and $W$ state [1], present quite different behaviors. For example, it
has been shown [2] that a three-qubit $W$ state is robust against losses of
qubits since it retains bipartite entanglement if any one qubit is traced
out, whereas a three-qubit GHZ state is fragile since the remaining two
partite states result in separable states. This feature makes $W$ states
very useful in various quantum information tasks. A $W$ state can be used as
a quantum channel for quantum key distribution [3], entangled-pairs
teleportation [4], quantum teleportation [5] and so on. During the past
years, many theoretical schemes have been proposed to generate the $W$ state
in many physical systems [6-18]. Moreover, the experimental demonstration of
$W$ states has been reported with up to eight trapped ions [19], four
optical modes [20], three capacitively-coupled superconducting phase qubits
[21], two superconducting phase qubits plus a resonant cavity [22], and
atomic ensembles in four quantum memories [23].

Instead of generating entangled states, we here focus on transferring quantum entanglement among qubits distributed in
different cavities. During the past years, much attention has been
paid to quantum entanglement transfer. For instances,
many proposals for transferring quantum entanglement via quantum teleportation protocols have been
presented [24-28], and schemes for transferring quantum entanglement based on cavity QED or circuit QED
have been also proposed [29-31]. Moreover, quantum entanglement transfer has been experimentally demonstrated in linear optics [32,33].

This work is also motivated for the following reason. Large-scale QIP will most likely need a large number of qubits,
and placing all of them in a single cavity may cause practical problems such as decreasing
the qubit-cavity coupling strength and increasing the cavity decay rate. Hence,
future QIP most likely requires quantum networks consisting of many cavities,
each hosting and coupled to multiple qubits. In this type of architecture, preparation, transfer, exchange, and manipulation of quantum
states (e.g., GHZ states, $W$ states and cluster states, etc.) will not only occur
among qubits in the same cavity, but also among qubits distributed in different cavities.

We consider a quantum system consisting of $2n$ cavities each hosting qubits.
The qubits can be made to be decoupled from their respective cavities before/after the
operation. And, the coupling of qubits with their cavities, which is
necessary for quantum operation, can be achieved, by prior adjustment of the
level spacings of the qubits or the frequencies of the cavities. In
the following, we will present a method to implement the coherent transfer
of a $W$-class entangled state from $n$ qubits in $n$ cavities onto $n$
qubits in another $n$ cavities. As shown below, this proposal has the
following advantages: (i) the entanglement transfer is performed without
excitation of the cavity photons, and thus decoherence induced by the cavity
decay is greatly suppressed during the entire operation; (ii) this proposal
needs only one coupler qubit and one operational step and does not require
using a classical pulse for the entanglement transfer, hence the engineering
complex is much reduced and the operation procedure is greatly simplified.
Finally, this proposal is quite general, and can be applied to accomplish
the same task with different types of qubits such as quantum dots, atoms, NV
centers, various superconducting qubits and so on. To the best of our knowledge,
how to transfer multipartite entanglement among qubits distributed in different cavities,
which are coupled by a single two-level qubit, has not been reported so far.

This paper is organized as follows. In Sec. II, we show a way to transfer a $%
n $-qubit $W$ state from $n$ qubits in $n$ cavities onto $n$ qubits in
another $n$ cavities. In Sec. III, as an example, we analyze the experimental
feasibility of transferring a three-qubit $W$ state in circuit QED. A
concluding summary is given in Sec. IV.

\begin{figure}[tbp]
\begin{center}
\includegraphics[bb=0 0 524 740, width=9.5 cm, clip]{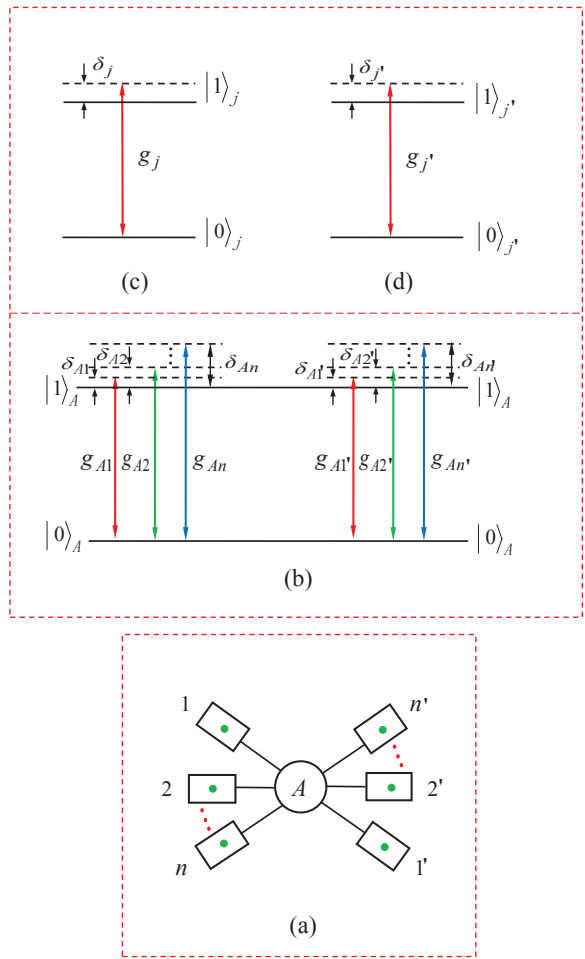} \vspace*{%
-0.08in}
\end{center}
\caption{(color online) (a) Diagram of a coupler qubit $A$ (a circle at the
center) and $2n$ cavities ($1,2,...,n,1^{\prime },2^{\prime },...,n^{\prime
} $) each hosting a qubit. A dark square represents a cavity while a green
dot labels a qubit placed in each cavity. (b) Dispersive interaction of the
coupler qubit $A$ with $2n$ cavities ($1,2,...,n,1^{\prime },2^{\prime
},...,n^{\prime }$). Cavity $j$ is coupled to qubit $A$ with coupling
constant $g_{Aj}$ and detuning $\protect\delta _{Aj}$ ($j=1,2,...,n$), and
Cavity $j^{\prime }$ is coupled to qubit $A$ with coupling constant $%
g_{Aj^{\prime }} $ and detuning $\protect\delta _{Aj^{\prime }}$ ($j^{\prime
}=1^{\prime },2^{\prime },...,n^{\prime }$). Here, $\protect\delta _{j}=%
\protect\delta _{Aj}=\protect\delta _{Aj^{\prime }}=\protect\delta %
_{j^{\prime }}$. (c) Dispersive interaction of qubit $j$ (placed in cavity $%
j $) with cavity $j$ ($j=1,2,...,n$). Here, $g_j$ is the coupling constant
while $\protect\delta _j$ is detuning. (d) Dispersive interaction of qubit $%
j^{\prime }$ (placed in cavity $j^{\prime }$) with cavity $j^{\prime }$ ($%
j^{\prime }=1^{\prime },2^{\prime },...,n^{\prime }$). Here, $g_{j^{\prime
}} $ is the coupling constant while $\protect\delta _{j^{\prime }}$ is
detuning.}
\label{fig:1}
\end{figure}

\section{W-STATE TRANSFER}

In the following, we first construct a Hamiltonian for the $W$ state
transfer and then describe the procedure for implementing the $W$-state
transfer.

\subsection{Hamiltonian}

Consider $n$ cavities $(1,2,...,n)$ and another $n$ cavities $(1^{\prime
},2^{\prime },...,n^{\prime }).$ The $2n$ cavities are connected by a
coupler qubit $A$, as illustrated in Fig.~1(a). The $n$ qubits placed in the
$n$ cavities $(1,2,...,n)$ are labelled as qubits $1,2,...,n$ while the $n$
qubits placed in the other $n$ cavities $(1^{\prime },2^{\prime
},...,n^{\prime })$ are denoted as qubits $1^{\prime },2^{\prime
},...,n^{\prime }.$ Assume that the coupling constant of qubit $j$ with
cavity $j$ is $g_j$ $(j=1,2,...,n)$ while the coupling constant of qubit $%
j^{\prime }$ with cavity $j^{\prime }$ is $g_{j^{\prime }}$ $(j^{\prime
}=1^{\prime },2^{\prime },...,n^{\prime }).$ The coupling and decoupling of
each qubit from its cavity (cavities) can be achieved by prior adjustment of
the qubit's level spacings. For superconducting devices, their level
spacings can be rapidly adjusted by varying external control parameters
(e.g., magnetic flux applied to phase, transmon, or flux qutrits; see, e.g.,
[34-36]).

Adjust the level spacings of the coupler qubit $A$ such that this qubit
interacts with the $2n$ cavities simultaneously. Denote $g_{Aj}$ as the
coupling constant of qubit $A$ with cavity $j$ while $g_{Aj^{\prime }}$ as
the coupling constant of qubit $A$ with cavity $j^{\prime }.$ In the
interaction picture under the free Hamiltonian of the whole system and
applying the rotating-wave approximation, we have
\begin{eqnarray}
H_{I} &=&\sum_{j=1}^{n}g_{j}\left( e^{i\delta _{j}t}a_{j}\sigma
_{j}^{+}+h.c.\right) +\sum_{j=1}^{n}g_{Aj}\left( e^{i\delta
_{Aj}t}a_{j}\sigma _{A}^{+}+h.c.\right)   \notag \\
&&+\sum_{j^{\prime }=1^{\prime }}^{n^{\prime }}g_{j^{\prime }}\left(
e^{i\delta _{j^{\prime }}t}a_{j^{\prime }}\sigma _{j^{\prime
}}^{+}+h.c.\right) +\sum_{j^{\prime }=1^{\prime }}^{n^{\prime
}}g_{Aj^{\prime }}\left( e^{i\delta _{Aj^{\prime }}t}a_{j^{\prime }}\sigma
_{A}^{+}+h.c.\right) ,
\end{eqnarray}%
where\ the first two terms correspond to the subsystem composed of the
coupler qubit $A,$ the $n$ cavities $(1,2,...,n)$ and the $n$ qubits $%
(1,2,...,n)$, while the last two terms correspond to the subsystem composed
of the coupler qubit $A,$ the $n$ cavities $(1^{\prime },2^{\prime
},...,n^{\prime })$ and the $n$ qubits $(1^{\prime },2^{\prime
},...,n^{\prime });$ $a_{j}$ ($a_{j^{\prime }}$) is the annihilation
operator for the mode of cavity $j$ ($j^{\prime }$); $\sigma
_{j}^{+}=\left\vert 1\right\rangle _{j}\left\langle 0\right\vert $ ($\sigma
_{j^{\prime }}^{+}=\left\vert 1\right\rangle _{j^{\prime }}\left\langle
0\right\vert $) is the raising operator of qubit $j$ ($j^{\prime }$); $%
\delta _{j},$ $\delta _{j^{\prime }},$ $\delta _{Aj},$ and $\delta
_{Aj^{\prime }}$ are the detunings, given by $\delta _{j}=\omega
_{10j}-\omega _{c_{j}},$ $\delta _{j^{\prime }}=\omega _{10j^{\prime
}}-\omega _{c_{j^{\prime }}},$ $\delta _{Aj}=\omega _{10A}-\omega _{c_{j}},$
and $\delta _{Aj^{\prime }}=\omega _{10A}-\omega _{c_{j^{\prime }}}$
[Fig.~1(b,c,d)].

In the case $\delta _{j}\gg g_{j,}\;\delta _{j^{\prime }}\gg g_{j^{\prime
},} $ $\delta _{Aj}\gg g_{Aj},\,$and $\delta _{Aj^{\prime }}\gg
g_{Aj^{\prime }}, $ there is no energy exchange between the qubit system and
the cavities. Under the condition of
\begin{eqnarray}
\frac{\left\vert \delta _{A(j+1)}-\delta _{Aj}\right\vert }{\delta
_{Aj}^{-1}+\delta _{A(j+1)}^{-1}} &\gg &g_{Aj}g_{A(j+1)},  \notag \\
\frac{\left\vert \delta _{A(j+1)^{\prime }}-\delta _{Aj^{\prime
}}\right\vert }{\delta _{Aj^{\prime }}^{-1}+\delta _{A(j+1)^{\prime }}^{-1}}
&\gg &g_{Aj^{\prime }}g_{A(j+1)^{\prime }},
\end{eqnarray}%
there is no interaction between the $n$ cavities $(1,2,...,n)$ and there is
no interaction between another $n$ cavities $(1^{\prime },2^{\prime
},...,n^{\prime }),$ which are induced by the coupler qubit $A$. For
simplicity, we set
\begin{equation}
\delta _{j}=\delta _{Aj}=\delta _{Aj^{\prime }}=\delta _{j^{\prime }} \text{
}(j=1,2,...,n).
\end{equation}%
Hence, we can obtain the following effective Hamiltonian [37,38]
\begin{eqnarray}
H_{\mathrm{eff}} &=&-\sum_{j=1}^{n}\frac{g_{j}^{2}}{\delta _{j}}\left(
\left\vert 0\right\rangle _{j}\left\langle 0\right\vert
a_{j}^{+}a_{j}-\left\vert 1\right\rangle _{j}\left\langle 1\right\vert
a_{j}a_{j}^{+}\right)  \notag \\
&&\ -\sum_{j=1}^{n}\frac{g_{Aj}^{2}}{\delta _{Aj}}\left( \left\vert
0\right\rangle _{A}\left\langle 0\right\vert a_{j}^{+}a_{j}-\left\vert
1\right\rangle _{A}\left\langle 1\right\vert a_{j}a_{j}^{+}\right)  \notag \\
&&\ -\sum_{j^{\prime }=1^{\prime }}^{n^{\prime }}\frac{g_{j^{\prime }}^{2}}{%
\delta _{j^{\prime }}}\left( \left\vert 0\right\rangle _{j^{\prime
}}\left\langle 0\right\vert a_{j^{\prime }}^{+}a_{j^{\prime }}-\left\vert
1\right\rangle _{j^{\prime }}\left\langle 1\right\vert a_{j^{\prime
}}a_{j^{\prime }}^{+}\right)  \notag \\
&&\ -\ \sum_{j^{\prime }=1^{\prime }}^{n^{\prime }}\frac{g_{Aj^{\prime }}^{2}%
}{\delta _{Aj^{\prime }}}\left( \left\vert 0\right\rangle _{A}\left\langle
0\right\vert a_{j^{\prime }}^{+}a_{j^{\prime }}-\left\vert 1\right\rangle
_{A}\left\langle 1\right\vert a_{j^{\prime }}a_{j^{\prime }}^{+}\right)
\notag \\
&&\ +\sum_{j=1}^{n}\lambda _{j}\left( \sigma _{j}^{+}\sigma _{A}+\sigma
_{j}\sigma _{A}^{+}\right)  \notag \\
&&\ +\sum_{j^{\prime }=1^{\prime }}^{n^{\prime }}\lambda _{j^{\prime
}}\left( \sigma _{j^{\prime }}^{+}\sigma _{A}+\sigma _{j^{\prime }}\sigma
_{A}^{+}\right)  \notag \\
&&\ +\sum_{j=1}^{n}\mu _{j}\left( a_{j}^{+}a_{j^{\prime }}+a_{j}a_{j^{\prime
}}^{+}\right) \left( \left\vert 1\right\rangle _{A}\left\langle 1\right\vert
-\left\vert 0\right\rangle _{A}\left\langle 0\right\vert \right) ,
\end{eqnarray}%
where $\lambda _{j}=g_{j}g_{Aj}/\delta _{j},$ $\lambda _{j^{\prime
}}=g_{j^{\prime }}g_{Aj^{\prime }}/\delta _{j},$ and $\mu
_{j}=g_{j}g_{j^{\prime }}/\delta _{j}$ because of the setting described by
Eq. (3). The last term of Eq. (4) describes the interaction between cavity $%
j $ and cavity $j^{\prime }$ ($j=1,2,...,n$), which is induced by the
coupler qubit $A.$

Assume that each cavity is initially in the vacuum state. The Hamiltonian
(4) then reduces to
\begin{equation}
H_{\mathrm{eff}}=H_{0}+H_{\mathrm{int}},
\end{equation}%
with
\begin{eqnarray}
H_{0} &=&\sum_{j=1}^{n}\frac{g_{j}^{2}}{\delta _{j}}\left\vert
1\right\rangle _{j}\left\langle 1\right\vert +\sum_{j=1}^{n}\frac{g_{Aj}^{2}%
}{\delta _{Aj}}\left\vert 1\right\rangle _{A}\left\langle 1\right\vert
\notag \\
&&\ +\sum_{j^{\prime }=1^{\prime }}^{n^{\prime }}\frac{g_{j^{\prime }}^{2}}{%
\delta _{j^{\prime }}}\left\vert 1\right\rangle _{j^{\prime }}\left\langle
1\right\vert +\sum_{j^{\prime }=1^{\prime }}^{n^{\prime }}\frac{%
g_{Aj^{\prime }}^{2}}{\delta _{Aj^{\prime }}}\left\vert 1\right\rangle
_{A}\left\langle 1\right\vert , \\
H_{\mathrm{int}} &=&\sum_{j=1}^{n}\lambda _{j}\left( \sigma _{j}^{+}\sigma
_{A}+\sigma _{j}\sigma _{A}^{+}\right) +\sum_{j^{\prime }=1^{\prime
}}^{n^{\prime }}\lambda _{j^{\prime }}\left( \sigma _{j^{\prime }}^{+}\sigma
_{A}+\sigma _{j^{\prime }}\sigma _{A}^{+}\right) .
\end{eqnarray}%
In a new interaction picture under the Hamiltonian $H_{0}$ and applying the
following conditions
\begin{equation}
\frac{g_{1}^{2}}{\delta _{1}}=\frac{g_{2}^{2}}{\delta _{2}}=\cdot \cdot
\cdot =\frac{g_{n}^{2}}{\delta _{n}}=\frac{g_{1^{\prime }}^{2}}{\delta
_{1^{\prime }}}=\frac{g_{2^{\prime }}^{2}}{\delta _{2^{\prime }}}=\cdot
\cdot \cdot =\frac{g_{n^{\prime }}^{2}}{\delta _{n^{\prime }}}=\chi
\end{equation}%
and

\begin{equation}
\frac{g_k^2}{\delta _k}=\frac{g_{k^{\prime }}^2}{\delta _{k^{\prime }}}%
=\sum_{j=1}^n\frac{g_{Aj}^2}{\delta _{Aj}}+\sum_{j^{\prime }=1^{\prime
}}^{n^{\prime }}\frac{g_{Aj^{\prime }}^2}{\delta _{Aj^{\prime }}}\;
\end{equation}
where $k\in \{1,2,...,n\}$ and $k^{\prime }\in \{1^{\prime },2^{\prime
},...,n^{\prime }\},$ we have

\begin{equation}
\widetilde{H}_{\mathrm{int}}=e^{iH_{0}t}H_{\mathrm{int}}e^{-iH_{0}t}=H_{%
\mathrm{int}},
\end{equation}%
where all phase factors, caused during the Hamiltonian transformation, are cancelled due to the use of conditions (8) and (9).

Eq.~(10) shows that the Hamiltonian $\widetilde{H}_{\mathrm{int}}$ takes the same form as the Hamiltonian $H_{\mathrm{int}}$ given in Eq.~(7). We set
\begin{eqnarray}
&\lambda _{1}=\lambda _{2}=...=\lambda _{n}=\lambda~,\nonumber\\
&\lambda _{1^{\prime}}=\lambda _{2^{\prime }}=...=\lambda _{n^{\prime }}=\lambda~.
\end{eqnarray}
In this case, the coupling constants $\lambda _{j}$ and $\lambda _{j^{\prime}}$ involved in the Hamiltonian $H_{\mathrm{int}}$ [see Eq. (7)] can be moved
out of the summation symbols. Thus, the Hamiltonian $\widetilde{H}_{\mathrm{int}}$ can be expressed
as
\begin{equation}
\widetilde{H}_{\mathrm{int}}=\lambda \left( J_{+}\sigma _{A}+J_{-}\sigma
_{A}^{+}\right) +\lambda \left( J_{+}^{\prime }\sigma _{A}+J_{-}^{\prime
}\sigma _{A}^{+}\right) ,
\end{equation}%
where $J_{+}=\sum_{j=1}^{n}\sigma _{j}^{+},J_{-}=\sum_{j=1}^{n}\sigma
_{j},J_{+}^{\prime }=\sum_{j^{\prime }=1^{\prime }}^{n^{\prime }}\sigma
_{j^{\prime }}^{+}$ and $J_{-}^{\prime }=\sum_{j^{\prime }=1^{\prime
}}^{n^{\prime }}\sigma _{j^{\prime }}.$ In the following, the Hamiltonian
(12) will be used to transfer the $W$ state from the $n$ qubits $\left(
1,2,...,n\right) $ to the other $n$ qubits $\left( 1^{\prime },2^{\prime
},...,n^{\prime }\right) .$

\subsection{$W$\textbf{-state transfer}}

The $W$ state $\left| W_{n-1,1}\right\rangle $ of $n$ qubits $(1,2,...,n)$
is described by [1]
\begin{equation}
\left| W_{n-1,1}\right\rangle =\frac 1{\sqrt{n}}\sum P_z\left|
0\right\rangle ^{\otimes \left( n-1\right) }\left| 1\right\rangle ,
\end{equation}
where $P_z$ is the symmetry permutation operator for qubits $(1,2,...,n),$ $%
\sum P_z\left| 0\right\rangle ^{\otimes \left( n-1\right) }\left|
1\right\rangle $ denotes the totally symmetric state in which $n-1$ of
qubits $(1,2,...,n)$ are in the state $\left| 0\right\rangle $ while the
remaining qubit is in the state $\left| 1\right\rangle .$ For instance, we
have $\left| W_{2,1}\right\rangle =\frac 1{\sqrt{3}}\left( \left|
001\right\rangle +\left| 010\right\rangle +\left| 100\right\rangle \right) $
when $n=3.$

Assume that (i) each cavity is initially in the vacuum state, (ii) the $n$
qubits $(1,2,...,n)$ are initially in the $W$ state $\left\vert
W_{n-1,1}\right\rangle $ described above, while the $n$ qubits $(1^{\prime
},2^{\prime },...,n^{\prime })$ are initially in the ground state, i.e.,
qubit $j^{\prime }$ is in the state $\left\vert 0\right\rangle _{j^{\prime
}} $, (iii) the coupler qubit $A$ is initially in the ground state $%
\left\vert 0\right\rangle _{A},$ and (iv)\ all qubits are decoupled from
their respective cavities.

To transfer the $W$ state, adjust the level spacings of each qubit
(including the coupler qubit $A$) to have the state of the qubit system
undergo the time evolution described by the Hamiltonian (12). Based on this
Hamiltonian and after returning to the original interaction picture by
performing a unitary transformation $e^{-iH_{0}t},$ it is easy to find that
the initial state $\left\vert W_{n-1,1}\right\rangle
_{12...n}\prod\limits_{j^{\prime }=1^{\prime }}^{n^{\prime }}\left\vert
0\right\rangle _{j^{\prime }}\left\vert 0\right\rangle _{A}$ of the qubit
system evolves into
\begin{eqnarray}
&&\ \ N\left[ e^{-i\chi t}\left( 1+\cos \Lambda t\right) \left\vert
W_{n-1,1}\right\rangle _{12...n}\prod\limits_{j^{\prime }=1^{\prime
}}^{n^{\prime }}\left\vert 0\right\rangle _{j^{\prime }}\left\vert
0\right\rangle _{A}\right.  \notag \\
&&\ \ +\left. e^{-i\chi t}\left( \cos \Lambda t-1\right)
\prod\limits_{j=1}^{n}\left\vert 0\right\rangle _{j}\left\vert
W_{n-1,1}\right\rangle _{1^{\prime }2^{\prime }...n^{\prime }}\left\vert
0\right\rangle _{A}\right]  \notag \\
&&\ \ \ \ \ \ \ -i\sqrt{N}e^{-i2\chi t}\sin \Lambda
t\prod\limits_{j=1}^{n}\left\vert 0\right\rangle _{j}\prod\limits_{j^{\prime
}=1^{\prime }}^{n^{\prime }}\left\vert 0\right\rangle _{j^{\prime
}}\left\vert 1\right\rangle _{A},
\end{eqnarray}%
where $N=1/2,$ and $\Lambda =\sqrt{2n}\left\vert \lambda \right\vert .$
Here, the factors $e^{-i\chi t}$ and $e^{-i2\chi t}$ were obtained by
performing the unitary transformation $e^{-iH_{0}t}$ and applying the
conditions (8) and (9).

One can see that for $t=\pi /\Lambda ,$ the state (14) becomes $%
\prod\limits_{j=1}^{n}\left\vert 0\right\rangle _{j}\left\vert
W_{n-1,1}\right\rangle _{1^{\prime }2^{\prime }...n^{\prime }}\left\vert
0\right\rangle _{A}$, which shows that the $n$ qubits ($1^{\prime
},2^{\prime },...,n^{\prime }$) are in the state $\left\vert
W_{n-1,1}\right\rangle $. Namely, the $W$ state of the qubits ($1,2,...,n$)
in $n$ cavities is transferred onto the $n$ qubits ($1^{\prime },2^{\prime
},...,n^{\prime }$) in the other $n$ cavities after the operation. To
maintain the $W$ state unaffected, the level spacings for each intracavity
qubit and the coupler qubit $A$ need to be adjusted back to the original
configuration after the above operation.

In above adjusting the qubit level spacings is unnecessary. Alternatively,
the coupling or decoupling of the qubits with the cavities can be obtained by
adjusting the frequency of each cavity. The rapid tuning of cavity
frequencies has been demonstrated in superconducting microwave cavities
(e.g., in less than a few nanoseconds for a superconducting transmission
line resonator [39]).

\subsection{Discussion}

For the approach to work, the following requirements need to be satisfied:

(i) The conditions (2), (3), (8), (9) and (11) need to be met for the
protocol to work. The condition (2) can be reached by prior design of
cavities with appropriate frequencies. The condition (3) is automatically
ensured for the identical qubits and pairs of cavities $j$ and $j^{\prime }$
with same frequency. Given\ $\delta _{1},\delta _{2},...,\delta _{n}$ and $%
\delta _{1^{\prime }},\delta _{2^{\prime }},...,\delta _{n^{\prime }},$ the
condition (8) can be met by adjusting the coupling constants $%
g_{1},g_{2},...,g_{n}$ and $g_{1}^{\prime },g_{2}^{\prime
},...,g_{n}^{\prime }$ (e.g., for solid-state qubits, the qubit-cavity
coupling constants can be readily changed by varying the positions of the
qubits embedded in their cavities). The condition (9) can be met by setting
\begin{equation}
g_{Aj}/g_{j}=g_{Aj^{\prime }}/g_{j^{\prime }}=1/\sqrt{2n},
\end{equation}%
where $j=1,2,...,n$ and $j^{\prime }=1^{\prime },2^{\prime },...,n^{\prime }$%
. Given\ $g_{j}$ and $g_{j^{\prime }}$, this requirement (15) can be
obtained by adjusting $g_{Aj}$ and $g_{Aj^{\prime }}.$ For a solid-state
coupler qubit $A$, $g_{Aj}$ and $g_{Aj^{\prime }}$ can be adjusted by
changing the qubit-cavity coupler capacitance $C_{j}$ and $C_{j^{\prime }},$
respectively (Fig. 2). Finally, note that the condition (11) is
automatically satisfied because of the conditions (3), (8), (9) and (15).
Overall, all necessary conditions here can be readily met.

(ii) The operation time required for the entanglement transfer needs to be
much shorter than $T_{1}$ (energy relaxation time) and $T_{2}$ (dephasing
time) of the level $\left\vert 1\right\rangle $, so that the decoherence,
caused by energy relaxation and dephasing of the qubits, is negligible for
the operation.

(iii) The lifetime of the cavity modes is given by
\begin{equation}
T_{\mathrm{cav}}=\frac{1}{2n}\min \{T_{\mathrm{cav}}^{1},T_{\mathrm{cav}%
}^{2},...,T_{\mathrm{cav}}^{n},T_{\mathrm{cav}}^{1^{\prime }},T_{\mathrm{cav}%
}^{2^{\prime }},...,T_{\mathrm{cav}}^{n^{\prime }}\},
\end{equation}
which needs to be much longer than the operation time$,$ such that the
effect of cavity decay is negligible for the operation.

(iv) When the coupler qubit $A$ is a solid-state qubit, there may exist an
inter-cavity cross coupling during the operation, which should be negligibly
small in order to reduce its effect on the operation fidelity. In the present
proposal, the unwanted inter-cavity crosstalk may not be a problem because
each cavity is virtually excited during the entire operation, as long as the
large detuning conditions can be well satisfied.

\section{POSSIBLE EXPERIMENTAL IMPLEMENTATION}

The physical systems composed of cavities and superconducting qubits have
been considered to be one of the most promising candidates for QIP [40-44].
In above a general type of qubit was considered. Let us now consider that
each qubit is a superconducting transmon qubit and each cavity is a
one-dimensional transmission line resonator (TLR). In addition, assume that
the coupler qubit $A$ is connected to each TLR via a capacitance. As an
example of experimental implementation, consider a setup in Fig. 2 for
transferring the three-qubit $W$ state from three transmon qubits ($1,2,3$)
each embedded in a different TLR to the other three transmon qubits ($1^{\prime },2^{\prime
},3^{\prime }$) each in another different TLR. To be more realistic, a third higher level $%
\left\vert 2\right\rangle $ for each qubit here needs to be considered
during the operations described above, since this level $\left\vert
2\right\rangle $ may be excited due to the $\left\vert 1\right\rangle
\leftrightarrow \left\vert 2\right\rangle $ transition induced by the cavity
mode(s), which will turn out to affect the operation fidelity. Therefore, to
quantify how well the proposed protocol works out, an analysis of the\
operation fidelity will be given for the $W$-state transfer, by taking this
higher level $\left\vert 2\right\rangle $ into account. Because of three
levels being considered, each qubit is renamed as a qutrit in the following.

\begin{figure}[tbp]
\begin{center}
\includegraphics[bb=152 437 533 694, width=7.5 cm, clip]{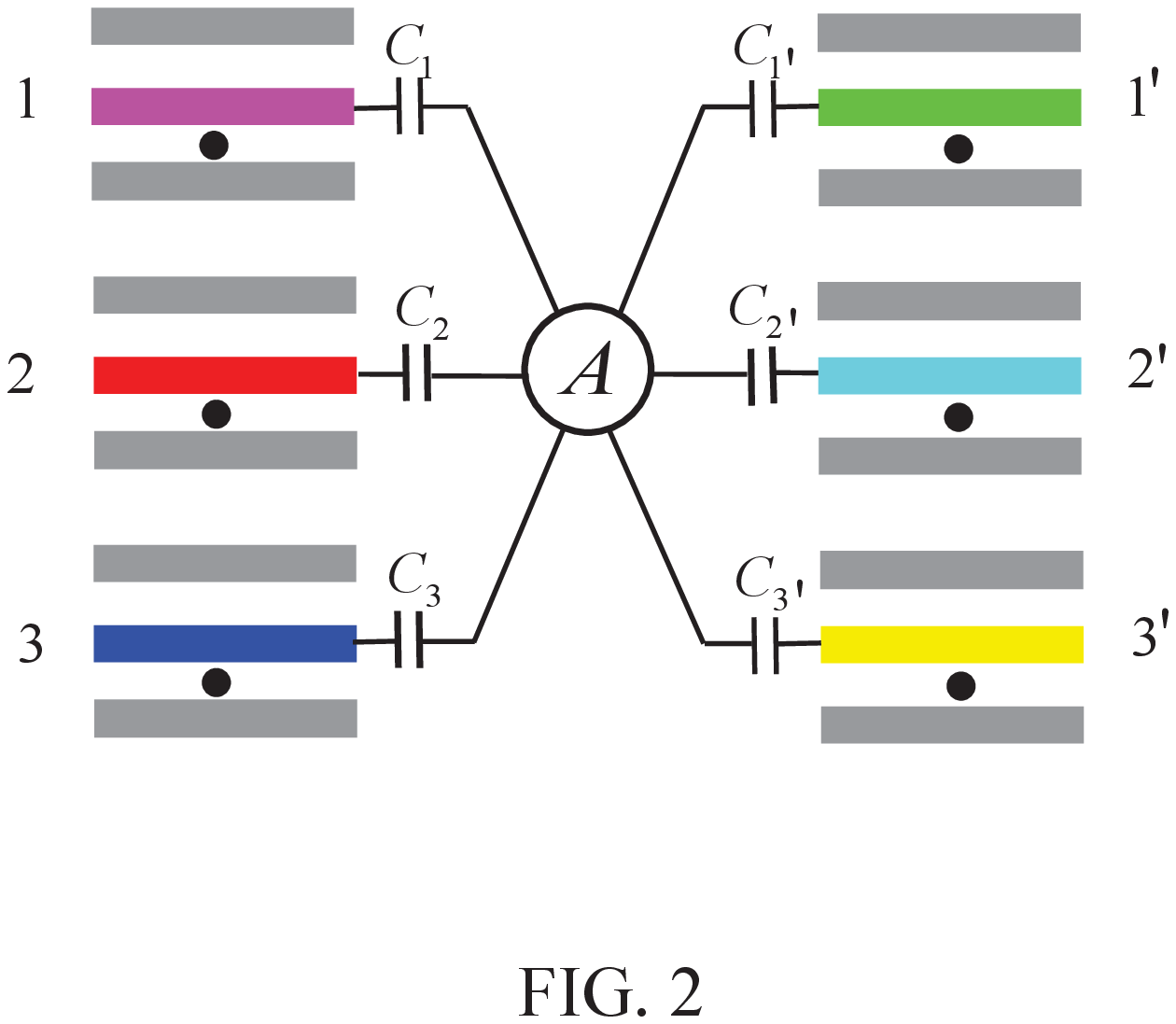} \vspace*{%
-0.08in}
\end{center}
\caption{(color online) Setup for six cavities ($1,2,3,1^{\prime },2^{\prime
},3^{\prime }$) coupled by a superconducting transmon qubit $A$. Each cavity
here is a one-dimensional coplanar waveguide transmission line resonator.
The circle $A$ represents a superconducting transmon qubit, which is
capacitively coupled to cavity j via a capacitance $C_j$ ($j = 1,2,3$) and
cavity $j^{\prime}$ via a capacitance $C_{j^{\prime}}$ ($j^{\prime} =
1^{\prime},2^{\prime},3^{\prime}$) . The six dark dots indicate the six
superconducting transmon qubits ($1,2,3,1^{\prime},2^{\prime},3^{\prime}$)
embedded in the six cavities, respectively. The interaction of qubits
(1,2,3) with their cavities are respectively illustrated in Fig.~3(a,b,c)
while the interaction of qubits ($1^{\prime},2^{\prime},3^{\prime}$) with
their cavities are respectively illustrated in Fig.~4(a,b,c). In addition,
the interaction of the coupler qubit $A$ with three cavities (1,2,3) is
illustrated in Fig.~3(d) while the interaction of the coupler qubit $A$ with
three cavities ($1^{\prime},2^{\prime},3^{\prime}$) is illustrated in
Fig.~4(d). Since three levels for each qubit is involved in our analysis,
each qubit is renamed a qutrit in Figs.~3 and 4.}
\label{fig:2}
\end{figure}

When the inter-cavity crosstalk coupling and the unwanted $\left|
1\right\rangle \leftrightarrow \left| 2\right\rangle $ transition of each
qutrit are considered, the Hamiltonian (1) is modified as follows

\begin{equation}
h_{I}=H_{I}+\Theta _{I},
\end{equation}%
where $H_{I}$ is the needed interaction Hamiltonian in Eq. (1) for $n=3$ and
$n^{\prime }=3^{\prime }$, while $\Theta _{I}$ is the unwanted interaction
Hamiltonian, given by
\begin{eqnarray}
\Theta _{I} &=&\sum_{j=1}^{3}\widetilde{g}_{j}\left( e^{i\widetilde{\delta }%
_{j}t}a_{j}\sigma _{21j}^{+}+h.c.\right) +\sum_{j=1}^{3}\widetilde{g}%
_{Aj}\left( e^{i\widetilde{\delta }_{Aj}t}a_{j}\sigma _{21A}^{+}+h.c.\right)
\notag \\
&&\ +\sum_{j^{\prime }=1^{\prime }}^{3^{\prime }}\widetilde{g}_{j^{\prime
}}\left( e^{i\widetilde{\delta }_{j^{\prime }}t}a_{j^{\prime }}\sigma
_{21j^{\prime }}^{+}+h.c.\right) +\sum_{j^{\prime }=1^{\prime }}^{3^{\prime
}}\widetilde{g}_{Aj^{\prime }}\left( e^{i\widetilde{\delta }_{Aj^{\prime
}}t}a_{j^{\prime }}\sigma _{21A}^{+}+h.c.\right)  \notag \\
&&\ +\sum_{k\neq l}g_{kl}\left( e^{-i\Delta
_{kl}t}a_{k}a_{l}^{+}+h.c.\right) ,
\end{eqnarray}%
where $k,l\in \{1,2,3,1^{\prime },2^{\prime },3^{\prime }\},\sigma
_{21j}^{+}=\left\vert 2\right\rangle _{j}\left\langle 1\right\vert ,\sigma
_{21j^{\prime }}^{+}=\left\vert 2\right\rangle _{j^{\prime }}\left\langle
1\right\vert ,$ and $\sigma _{21A}^{+}=\left\vert 2\right\rangle
_{A}\left\langle 1\right\vert .$ The first term represents the unwanted
off-resonant coupling between the mode of cavity $j$ and the $\left\vert
1\right\rangle \leftrightarrow \left\vert 2\right\rangle $ transition of
qutrit $j$, with coupling constant $\widetilde{g}_{j}$ and detuning $%
\widetilde{\delta }_{j}=\omega _{21j}-\omega _{c_{j}}$ [Fig.~3(a,b,c)],
while the second term indicates the unwanted off-resonant coupling between
the mode of cavity $j$ and the $\left\vert 1\right\rangle \leftrightarrow
\left\vert 2\right\rangle $ transition of qutrit $A$, with coupling constant
$\widetilde{g}_{Aj}$ and detuning $\widetilde{\delta }_{Aj}=\omega
_{21A}-\omega _{c_{j}}$ [Fig.~3(d)]. The third term represents the unwanted
off-resonant coupling between the mode of cavity $j^{\prime }$ and the $%
\left\vert 1\right\rangle \leftrightarrow \left\vert 2\right\rangle $
transition of qutrit $j^{\prime }$, with coupling constant $\widetilde{g}%
_{j^{\prime }}$ and detuning $\widetilde{\delta }_{j^{\prime }}=\omega
_{21j^{\prime }}-\omega _{c_{j^{\prime }}}$ [Fig.~4(a,b,c)], while the
fourth term indicates the unwanted off-resonant coupling between the mode of
cavity $j^{\prime }$ and the $\left\vert 1\right\rangle \leftrightarrow
\left\vert 2\right\rangle $ transition of qutrit $A$, with coupling constant
$\widetilde{g}_{Aj^{\prime }}$ and detuning $\widetilde{\delta }_{Aj^{\prime
}}=\omega _{21A}-\omega _{c_{j^{\prime }}}$ [Fig.~4(d)]. The last term
describes the inter-cavity crosstalk between any two cavities, with $\Delta
_{kl}=\omega _{c_{k}}-\omega _{c_{l}}=\delta _{l}-\delta _{k}$ (the
frequency difference between two cavities $k$ and $l$) and $g_{kl}$ (the
inter-cavity coupling constant between two cavities $k$ and $l$).

\begin{figure}[tbp]
\begin{center}
\includegraphics[bb=5 106 591 777, width=12.5 cm, clip]{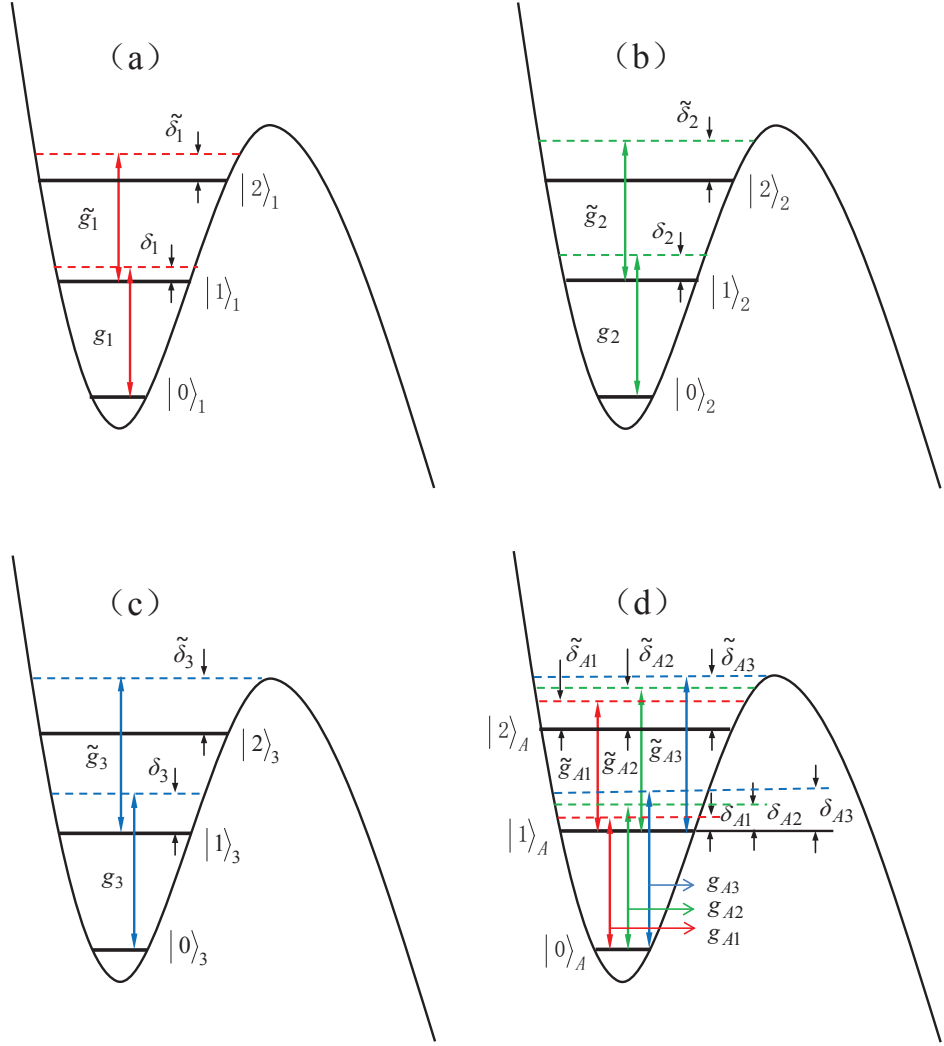} \vspace*{%
-0.08in}
\end{center}
\caption{(Color online) Illustration of the interaction between qutrits
(1,2,3,$A$) and three cavities (1,2,3). (a) Cavity $1$ is dispersively
coupled to the $\left\vert 0\right\rangle \leftrightarrow \left\vert
1\right\rangle $ transition with coupling constant $g_{1}$ and detuning $%
\protect\delta _{1},$ but far-off resonant (i.e., more detuned) with the $%
\left\vert 1\right\rangle \leftrightarrow \left\vert 2\right\rangle $
transition of qutrit $1$ with coupling constant $\widetilde{g}_{1}$ and
detuning $\widetilde{\protect\delta }_{1}$. (b) [and (c)] corresponds to the
case that cavity $2$ ($3$) is dispersively coupled to the $\left\vert
0\right\rangle \leftrightarrow \left\vert 1\right\rangle $ transition but
far-off resonant with the $\left\vert 1\right\rangle \leftrightarrow
\left\vert 2\right\rangle $ transition of qutrit $2$ ($3$). (d) Cavities ($%
1,2,3$) dispersively interact with the $\left\vert 0\right\rangle
\leftrightarrow \left\vert 1\right\rangle $ transition of qutrit $A$ with
coupling constants ($g_{A1},g_{A2},g_{A3}$) and detunings ($\protect\delta %
_{A1},\protect\delta _{A2},\protect\delta _{A3}$), respectively; but they
are far-off resonant with the $\left\vert 1\right\rangle \leftrightarrow
\left\vert 2\right\rangle $ transition of qutrit $A$ with coupling constants (%
$\widetilde{g}_{A1},\widetilde{g}_{A2},\widetilde{g}_{A3}$) and detunings ($%
\widetilde{\protect\delta }_{A1},\widetilde{\protect\delta }_{A2},\widetilde{%
\protect\delta }_{A3}$), respectively. Here, $\protect\delta _{j}=\protect%
\omega _{10j}-\protect\omega _{cj},\widetilde{\protect\delta }_{j}=\protect%
\omega _{21j}-\protect\omega _{cj},\protect\delta _{Aj}=\protect\omega %
_{10A}-\protect\omega _{cj},$ and $\widetilde{\protect\delta }_{Aj}=\protect%
\omega _{21A}-\protect\omega _{cj}$ ($j=1,2,3$), where $\protect\omega %
_{10j} $ ($\protect\omega _{21j}$) is the $\left\vert 0\right\rangle
\leftrightarrow \left\vert 1\right\rangle $ ($\left\vert 1\right\rangle
\leftrightarrow \left\vert 2\right\rangle $) transition frequency of qutrit $%
j$, $\protect\omega _{10A}$ ($\protect\omega _{21A}$) is the $\left\vert
0\right\rangle \leftrightarrow \left\vert 1\right\rangle $ ($\left\vert
1\right\rangle \leftrightarrow \left\vert 2\right\rangle $) transition
frequency of qutrit $A$, and $\protect\omega _{cj}$ is the frequency of
cavity $j$.}
\label{fig:3}
\end{figure}

\begin{figure}[tbp]
\begin{center}
\includegraphics[bb=5 113 593 775, width=12.5 cm, clip]{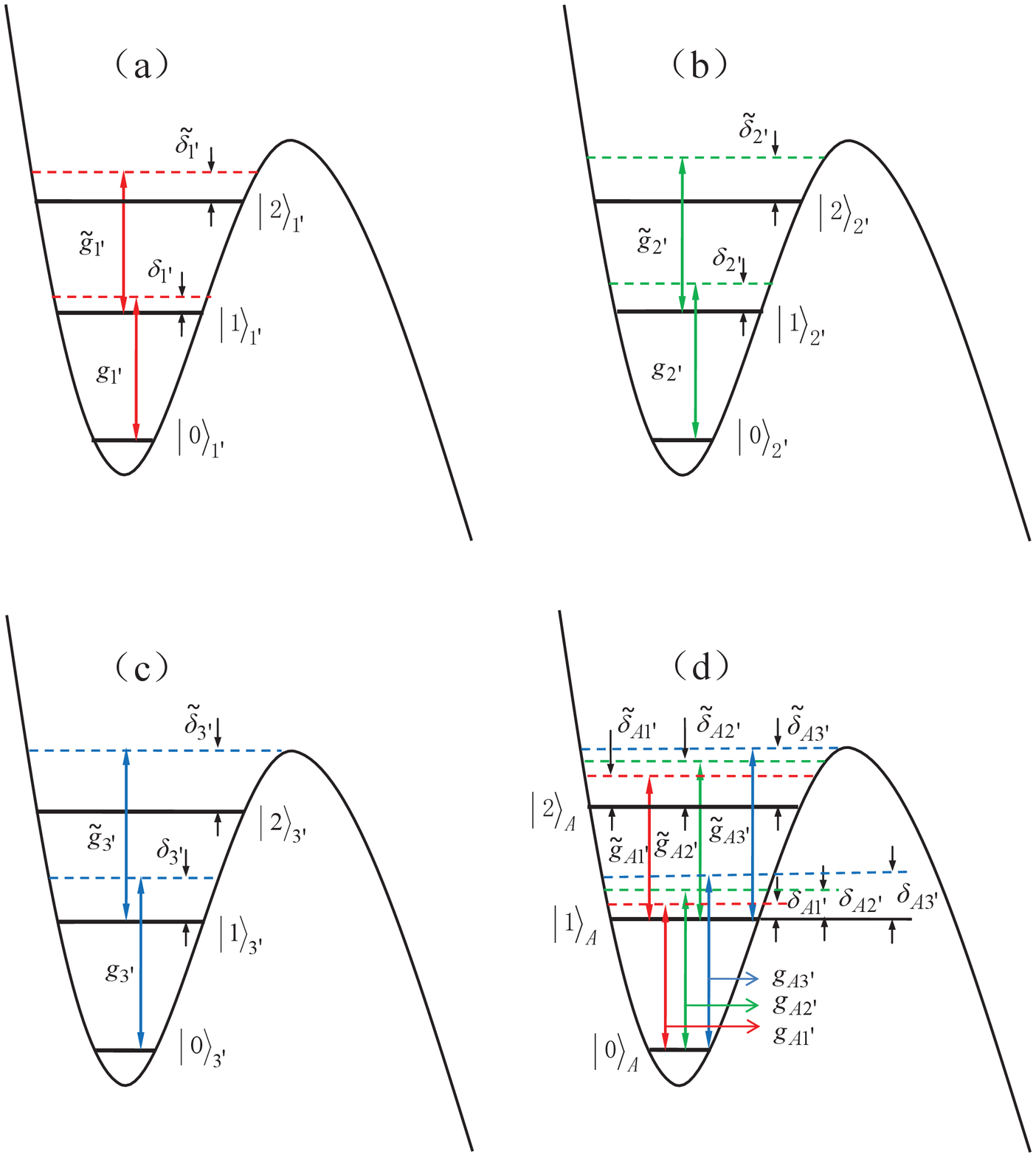} \vspace*{%
-0.08in}
\end{center}
\caption{(Color online) Illustration of the interaction between qutrits ($%
1^{\prime},2^{\prime},3^{\prime},A$) and three cavities ($%
1^{\prime},2^{\prime},3^{\prime}$). (a) Cavity $1^{\prime}$ is dispersively
coupled to the $\left\vert 0\right\rangle \leftrightarrow \left\vert
1\right\rangle $ transition with coupling constant $g_{1^{\prime}}$ and
detuning $\protect\delta _{1^{\prime}},$ but far-off resonant (i.e., more
detuned) with the $\left\vert 1\right\rangle \leftrightarrow \left\vert
2\right\rangle $ transition of qutrit $1^{\prime}$ with coupling constant $%
\widetilde{g}_{1^{\prime}}$ and detuning $\widetilde{\protect\delta }%
_{1^{\prime}}$. (b) [and (c)] corresponds to the case that cavity $%
2^{\prime} $ ($3^{\prime}$) is dispersively coupled to the $\left\vert
0\right\rangle \leftrightarrow \left\vert 1\right\rangle $ transition but
far-off resonant with the $\left\vert 1\right\rangle \leftrightarrow
\left\vert 2\right\rangle $ transition of qutrit $2^{\prime}$ ($3^{\prime}$%
). (d) Cavities ($1^{\prime},2^{\prime},3^{\prime}$) dispersively interact
with the $\left\vert 0\right\rangle \leftrightarrow \left\vert
1\right\rangle $ transition of qutrit $A$ with coupling constants ($%
g_{A1^{\prime}},g_{A2^{\prime}},g_{A3^{\prime}}$) and detunings ($\protect%
\delta _{A1^{\prime}},\protect\delta _{A2^{\prime}},\protect\delta %
_{A3^{\prime}}$), respectively; but they are far-off resonant with the $%
\left\vert 1\right\rangle \leftrightarrow \left\vert 2\right\rangle $
transition of qutrit $A$ with coupling constants ($\widetilde{g}%
_{A1^{\prime}},\widetilde{g}_{A2^{\prime}},\widetilde{g}_{A3^{\prime}}$) and
detunings ($\widetilde{\protect\delta }_{A1^{\prime}},\widetilde{\protect%
\delta }_{A2^{\prime}},\widetilde{\protect\delta }_{A3^{\prime}}$),
respectively. Here, $\protect\delta _{j^{\prime}}=\protect\omega %
_{10j^{\prime}}-\protect\omega _{cj^{\prime}},\widetilde{\protect\delta }%
_{j^{\prime}}=\protect\omega _{21j^{\prime}}-\protect\omega _{cj^{\prime}},%
\protect\delta _{Aj^{\prime}}=\protect\omega _{10A}-\protect\omega %
_{cj^{\prime}},$ and $\widetilde{\protect\delta }_{Aj^{\prime}}=\protect%
\omega _{21A}-\protect\omega _{cj^{\prime}}$ ($j^{\prime}=1^{\prime},2^{%
\prime},3^{\prime}$), where $\protect\omega _{10j^{\prime}} $ ($\protect%
\omega _{21j^{\prime}}$) is the $\left\vert 0\right\rangle \leftrightarrow
\left\vert 1\right\rangle $ ($\left\vert 1\right\rangle \leftrightarrow
\left\vert 2\right\rangle $) transition frequency of qutrit $j^{\prime}$,
and $\protect\omega _{cj^{\prime}}$ is the frequency of cavity $j^{\prime}$.}
\label{fig:4}
\end{figure}

The dynamics of the lossy system, with finite qutrit relaxation and
dephasing and photon lifetime included, is determined by the following
master equation

\begin{eqnarray}
\frac{d\rho }{dt} &=&-i\left[ h_I,\rho \right] +\sum_{j=1}^3\kappa _j%
\mathcal{L}\left[ a_j\right] +\sum_{j^{\prime }=1^{\prime }}^3\kappa
_{j^{\prime }}\mathcal{L}\left[ a_{j^{\prime }}\right]  \notag \\
&&+\sum_l\left\{ \gamma _l\mathcal{L}\left[ \sigma _l^{-}\right] +\gamma
_{21l}\mathcal{L}\left[ \sigma _{21l}^{-}\right] +\gamma _{20l}\mathcal{L}%
\left[ \sigma _{20l}^{-}\right] \right\}  \notag \\
&&+\sum_l\left\{ \gamma _{l,\varphi 1}\left( \sigma _{11l}\rho \sigma
_{11l}-\sigma _{11l}\rho /2-\rho \sigma _{11l}/2\right) \right\}  \notag \\
&&+\sum_l\left\{ \gamma _{l,\varphi 2}\left( \sigma _{22l}\rho \sigma
_{22l}-\sigma _{22l}\rho /2-\rho \sigma _{22l}/2\right) \right\}
\end{eqnarray}
where $l\in \{1,2,3,1^{\prime },2^{\prime },3^{\prime },A\},\sigma
_{20l}^{-}=\left| 0\right\rangle _l\left\langle 2\right| ,\sigma
_{11l}=\left| 1\right\rangle _l\left\langle 1\right| ,\sigma _{22l}=\left|
2\right\rangle _l\left\langle 2\right| ;$ and $\mathcal{L}\left[ \Lambda %
\right] =\Lambda \rho \Lambda ^{+}-\Lambda ^{+}\Lambda \rho /2-\rho \Lambda
^{+}\Lambda /2,$ with $\Lambda =a_j,a_{j^{\prime }},\sigma _l^{-},\sigma
_{21l}^{-},\sigma _{20l}^{-}.$ Here, $\kappa _j$ is the photon decay rate of
cavity $a_j$ while $\kappa _{j^{\prime }}$ is the photon decay rate of
cavity $a_{j^{\prime }}$. In addition, $\gamma _l$ is the energy relaxation
rate of the level $\left| 1\right\rangle $ of qutrit $l$, $\gamma _{21l}$ ($%
\gamma _{20l}$) is the energy relaxation rate of the level $\left|
2\right\rangle $ of qutrit $l$ for the decay path $\left| 2\right\rangle
\rightarrow \left| 1\right\rangle $ ($\left| 0\right\rangle $), and $\gamma
_{l,\varphi 1}$ ($\gamma _{l,\varphi 2}$) is the dephasing rate of the level
$\left| 1\right\rangle $ ($\left| 2\right\rangle $) of qutrit $l.$

The fidelity of the operation is given by
\begin{equation}
\mathcal{F}=\sqrt{\left\langle \psi _{\mathrm{id}}\right| \rho \left| \psi _{%
\mathrm{id}}\right\rangle} ,
\end{equation}
where $\left| \psi _{\mathrm{id}}\right\rangle $ is the output state $%
\prod_{j=1}^3\left| 0\right\rangle _j\left| W_{2,1}\right\rangle
_{1^{\prime}2^{\prime}3^{\prime}}\left| 0\right\rangle _A\prod_{j=1}^3\left|
0\right\rangle _{c_j}\prod_{j^{\prime }=1^{\prime }}^{3^{\prime }}\left|
0\right\rangle _{c_{j^{\prime }}}$ of an ideal system (i.e., without
dissipation, dephasing, and crosstalk) as discussed in the previous
section; and $\rho$ is the final density operator of the system when the
operation is performed in a realistic physical \textrm{system}.

Without loss of generality, consider six identical superconducting transmon
qutrits. According to the condition (3), set $\delta _{1}=\delta
_{A1}=\delta _{A1^{\prime }}=\delta _{1^{\prime }}=-2\pi \times 0.5$ GHz, $%
\delta _{2}=\delta _{A2}=\delta _{A2^{\prime }}=\delta _{2^{\prime }}=-2\pi
\times 1.0$ GHz, $\delta _{3}=\delta _{A3}=\delta _{A3^{\prime }}=\delta
_{3^{\prime }}=-2\pi \times 1.5$ GHz. Set $\widetilde{\delta }_{j}=\delta
_{j}-2\pi \times 400$ MHz$,$ $\widetilde{\delta }_{Aj}=\delta _{Aj}-2\pi
\times 400$ MHz, $\widetilde{\delta }_{j^{\prime }}=\delta _{j^{\prime
}}-2\pi \times 400$ MHz, and $\widetilde{\delta }_{Aj^{\prime }}=\delta
_{Aj^{\prime }}-2\pi \times 400$ MHz (an anharmonicity readily achieved in
experiments [45]). For transmon qutrits, the typical transition frequency
between two neighbor levels is between 4 and 10 GHz. Thus, choose $\omega
_{10A},\omega _{10j},\omega _{10j}\sim 2\pi \times 6.5$ GHz. Given $\{\delta
_{1},\delta _{2},\delta _{3},\delta _{1^{\prime }},\delta _{2^{\prime
}},\delta _{3^{\prime }},$ $g_{1}\}$, the coupling constants $g_{2},$ $g_{3},
$ $g_{2^{\prime }\text{, }}$and $g_{3^{\prime }}$ are determined based on
Eq. (8). In addition, $g_{Aj}$ and $g_{Aj^{\prime }}$ are determined by Eq.
(15)$,$ given $g_{j}$ and $g_{j^{\prime }}$ ($j=1,2,3;j^{\prime }=1^{\prime
},2^{\prime },3^{\prime }$). For the present case, $n=3.$ Next, one has $%
\widetilde{g}_{j}\sim \sqrt{2}g_{j},$ $\widetilde{g}_{j^{\prime }}\sim \sqrt{%
2}g_{j^{\prime }},$ $\widetilde{g}_{Aj}\sim \sqrt{2}g_{Aj},$ and $\widetilde{%
g}_{Aj^{\prime }}\sim \sqrt{2}g_{Aj^{\prime }}$ for the transmon qutrit
here. Choose $\kappa _{j}^{-1}=\kappa _{j^{\prime }}^{-1}=5$ $\mu $s, $%
\gamma _{l,\varphi 1}^{-1}=\gamma _{l,\varphi 2}^{-1}=5$ $\mu $s, $\gamma
_{l}^{-1}=10$ $\mu $s, $\gamma _{21l}^{-1}=5$ $\mu $s, and $\gamma
_{20l}^{-1}=25$ $\mu $s (a conservative consideration), which are available
in experiment because $T_{1}$ and $T_{2}$ can be made to be on the order of $%
20-60$ $\mu $s for state-of-the-art superconducting transmon devices at the
present time [46-48]. Note that for a transmon qutrit with the three levels
considered here, the $\left\vert 0\right\rangle \leftrightarrow \left\vert
2\right\rangle $ dipole matrix element is much smaller than that of the $%
\left\vert 0\right\rangle \leftrightarrow \left\vert 1\right\rangle $ and $%
\left\vert 1\right\rangle \leftrightarrow \left\vert 2\right\rangle $
transitions. Thus, $\gamma _{20l}^{-1}\gg \gamma _{l}^{-1},\gamma
_{21l}^{-1}.$

\begin{figure}[tbp]
\begin{center}
\includegraphics[bb=0 0 800 500, width=10.5 cm, clip]{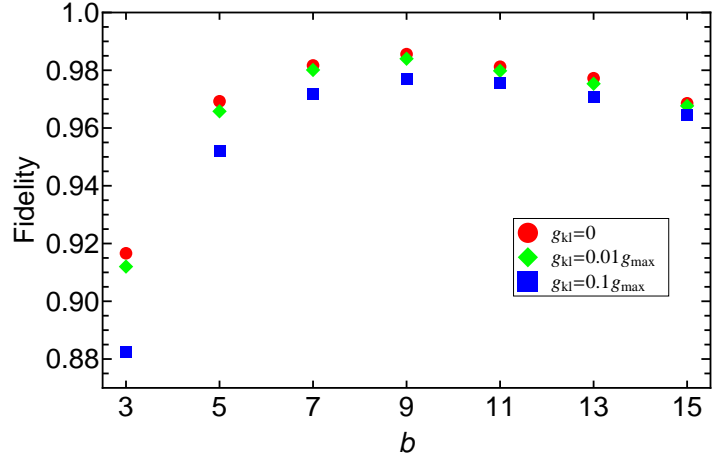} \vspace*{%
-0.08in}
\end{center}
\caption{(Color online) Fidelity of the $W$-state transfer versus the
normalized detuning $b=\left\vert \protect\delta _{1}\right\vert /g_{1}=%
\protect\delta _{1^{\prime}}/g_{1^{\prime}}$. Refer to the text for the
parameters used in the numerical calculation. Here, $g_{kl}$ are the
coupling strengths between cavities $k$ and $l$ ( $k\neq l;\,k,l\in \left\{
1,2,3,1^{\prime},2^{\prime},3^{\prime}\right\} $), which are taken to be the
same for simplicity.}
\label{fig:5}
\end{figure}

According to Eqs. (3) and (8), it is easy to see $g_{j}=g_{j^{\prime }}$ ($%
j=1,2,...,n$). For the parameters chosen above, the fidelity versus $%
b=\left\vert \delta _{1}\right\vert /g_{1}=\left\vert \delta _{1^{\prime
}}\right\vert /g_{1^{\prime }}$ is plotted in Fig.~5 for $g_{kl}=0,\
0.01g_{\max },$ and $0.1g_{\max }$, where $g_{\max }=\max
\{g_{A1},g_{A2},g_{A3},g_{A1^{\prime }},\\g_{A2^{\prime }},g_{A3^{\prime }}\}.$
Fig.~5 shows that for $g_{kl}\leq 0.01g_{\max }$, the effect of intercavity
cross coupling between the cavities on the fidelity of the operation is
negligible, which can be seen by comparing the top two curves. It can be
seen from Fig.~5 that for $b\sim 9$ and $g_{kl}=0.01g_{\max }\ (0.1g_{\max })
$, a high fidelity $\sim 98.4\%$ ($97.7\%$) is available.
For b=9, the operational time is only $0.081\mu s$, which is much less than decoherence and dephasing times of the system.
Moreover, the time averaged photon number in each cavity is $\sim0.006$,
which means the assumption of no excitation of cavity photons can be guaranteed safely.

The condition $g_{kl}\leq 0.01g_{\max }$ is not difficult to satisfy with
the typical capacitive cavity-qutrit coupling illustrated in Fig.~2. As
discussed in [49], as long as the cavities are physically well separated, the
inter-cavity cross-talk coupling strength is $g_{kl}\sim
g_{Ak}C_{l}/C_{\Sigma },g_{Al}C_{k}/C_{\Sigma },$ where\ $C_{\Sigma
}=\sum_{j=1}^{3}C_{j}+\sum_{j^{\prime }=1^{\prime }}^{3^{\prime
}}C_{j^{\prime }}+C_{q}$ with the qutrit's self capacitance $C_{q}.$  For $%
C_{1},C_{2},C_{3},C_{1^{\prime }},C_{2^{\prime }},C_{3^{\prime }}\sim 1$ fF
and $C_{\Sigma }\sim 10^{2}$ fF (the typical values in experiment [39]), one
has $g_{kl}\sim 0.01g_{Ak},0.01g_{Al}$ ($k,l\in \{1,2,3,1^{\prime
},2^{\prime },3^{\prime }$). Because of $g_{A1,}$ $g_{A2,}$ $%
g_{A3},g_{A1^{\prime }},g_{A2^{\prime }},g_{A3^{\prime }}\leq g_{\max },$
the condition $g_{kl}\leq 0.01g_{\max }$ can be readily met.

For $b\sim 9$, the coupling strengths are $%
\{g_{1},g_{2},g_{3},g_{A1},g_{A2},g_{A3}\}\sim $ $%
\{55.6,78.6,96.2, \\ 22.7,32.1,39.3\}$ MHz . The same values apply to $%
\{g_{1^{\prime }},g_{2^{\prime }},g_{3^{\prime }},g_{A1^{\prime
}},g_{A2^{\prime }},g_{A3^{\prime }}\},$ respectively. Note that the
coupling strengths with this value are readily achievable in experiment
because $g/\left( 2\pi \right) \sim 360$\textbf{\ }MHz has been reported for
a superconducting transmon qubit coupled to a one-dimensional standing-wave
CPW (coplanar waveguide) resonator [50,51]. For the transmon qutrits with
frequency $\omega _{10}/\left( 2\pi \right) \sim 6.5$ GHz chosen above, we
have $\omega _{c1}/2\pi ,\omega _{c1^{\prime }}/2\pi \sim 7.0$ GHz, $\omega
_{c2}/2\pi ,\omega _{c2^{\prime }}/2\pi \sim $ $7.5$ GHz, and $\omega
_{c3}/2\pi ,\omega _{c3^{\prime }}/2\pi \sim $ $8.0$ GHZ. For these cavity
frequencies and the values of $\kappa _{j}^{-1}$ and $\kappa _{j^{\prime
}}^{-1}$ used in the numerical calculation, the required quality factors for
the six cavities are $Q_{1},Q_{1^{\prime }}\sim 2.2\times 10^{5},$ $%
Q_{2},Q_{2^{\prime }}\sim 2.4\times 10^{5},$ and $Q_{3,3^{\prime }}\sim
2.5\times 10^{5},$ respectively. It should be mentioned that superconducting
CPW resonators with a loaded quality factor $Q\sim 10^{6}$ have been
experimentally demonstrated [52,53]. The analysis given here demonstrates
that high-fidelity transfer of the three-qubit $W$ state by using this
proposal is feasible within present-day circuit QED technique. It should be
remarked that further investigation is needed for each particular
experimental setup. However, it requires a rather lengthy and complex
analysis, which is beyond the scope of this theoretical work.

It is necessary to test whether high-fidelity transfer of the $W$ state can still be obtained
if conditions (3), (8), (9) and (11) are not fully satisfied. We assume that Eq. (3) is broken as:
\begin{equation}
\delta _{j}=\delta _{Aj}\neq\delta _{Aj^{\prime }}=\delta _{j^{\prime }} \text{
}(j=1,2,3).
\end{equation}
We set $\delta _{1}=\delta
_{A1}=-2\pi \times 0.5$ GHz, $\delta _{2}=\delta _{A2}=-2\pi
\times 1.0$ GHz, $\delta _{3}=\delta _{A3}=-2\pi \times 1.5$ GHz,
but set $\delta_{Aj^{\prime}}=\delta_{j^{\prime}}=r \delta_j$ ($j=1,2,3$), where $r$ is a new parameter describing
the degree of breakage. We adopt the previous values of $g_j, g_{j'}, g_{Aj}, g_{Aj'}$ used in Fig. 5.
For $r\neq1$, the values taken by $\delta_{Aj^{\prime}}$ and $\delta_{j^{\prime}}$ are not equal to the previous ones used in Fig. 5.
Thus, it is obvious that the conditions given in Eqs. (3), (8), (9) and (11) are broken simultaneously.
Fig.~6 shows the change of fidelity versus $r$, which is plotted for $b=9$ and $g_{kl}=0.01g_{max}$.
From Fig.~6, one can see that a high fidelity $\mathcal{F}\gtrsim0.969$ can be maintained for $0.9<r<1.1$.

\begin{figure}[tbp]
\begin{center}
\includegraphics[bb=0 0 600 390, width=10.5 cm, clip]{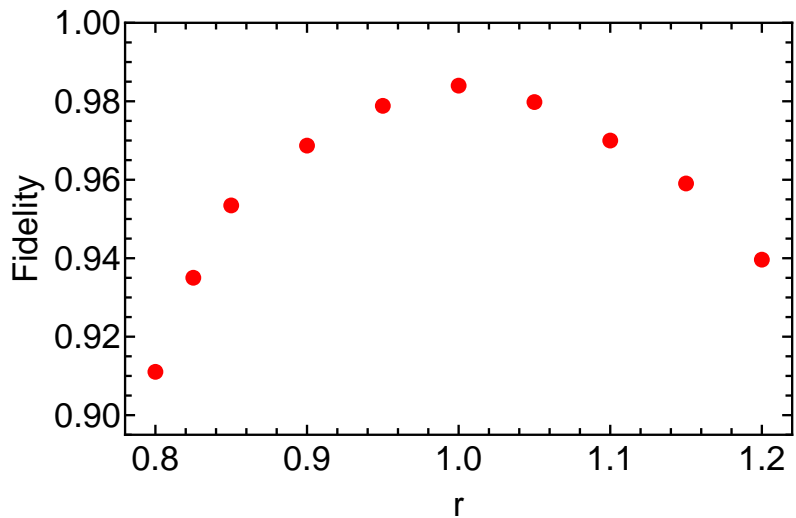} \vspace*{%
-0.08in}
\end{center}
\caption{(Color online) Fidelity of the $W$-state transfer versus the
ratio $r=\delta_{j'}/\delta_j$, plotted for $b=9$ and $g_{kl}=0.01g_{max}$.}
\label{fig:6}
\end{figure}

\section{CONCLUSION}

We have shown that transferring the $W$-class entangled states of multiple
qubits among different cavities can be realized by using a single
coupler qubit. As shown above, this proposal offers some advantages and
features. The entanglement transfer does not employ cavity photons as
quantum buses, thus decoherence caused due to the cavity decay is greatly
suppressed during the operation. Only one coupler qubit is needed to connect
with all cavities such that the circuit complex is greatly reduced.
Moreover, only one step of operation is required and no classical pulse is
need, so that the operation is much simplified. The numerical simulation
shows that high-fidelity transfer of the three-qubit $W$ state is feasible
for the current circuit QED technology. The method presented here is quite
general, and can be applied to accomplish the same task with different types
of qubits such as quantum dots, superconducting qubits (e.g., phase, flux
and charge qubits), NV centers, and atoms.

\section*{Acknowledgment}

C.P.Y. was supported in part by the National Natural Science Foundation of
China under Grant Nos. 11074062 and 11374083, the Zhejiang Natural Science
Foundation under Grant No. LZ13A040002, and the funds from Hangzhou Normal
University under Grant Nos. HSQK0081 and PD13002004. Q.P.S. was supported in part by the National
Natural Science Foundation of China under Grant No. 11504075, 11247008 and the Zhejiang Natural Science
Foundation under Grant No. LQ12A05004. This work was also
supported by the funds of Hangzhou City for the Hangzhou-City Quantum
Information and Quantum Optics Innovation Research Team.\\
\textbf{Conflict of Interest}: The authors declare that they have no conflict of interest.

\end{document}